# Detect adverse drug reactions for drug Atorvastatin

Yihui Liu[1]

[1]Institute of Intelligent Information Processing
Shandong Polytechnic University, China
Yihui_liu_2005@yahoo.co.uk

Uwe Aickelin[2]

[2]Department of Computer Science, University of Nottingham, UK

*Abstract*—Adverse drug reaction (ADR) is widely concerned for public health issue. In this study we propose an original approach to detect the ADRs using feature matrix and feature selection. The experiments are carried out on the drug Atorvastatin. Major side effects for the drug are detected and better performance is achieved compared to other computerized methods. The detected ADRs are based on the computerized method, further investigation is needed.

## I. INTRODUCTION

Adverse drug reaction (ADR) is widely concerned for public health issue. ADRs are one of most common causes to withdraw some drugs from market [1]. Now two major methods for detecting ADRs are spontaneous reporting system (SRS) [2, 3], and prescription event monitoring (PEM) [4, 5]. The World Health Organization (WHO) defines a signal in pharmacovigilance as "any reported information on a possible causal relationship between an adverse event and a drug, the relationship being unknown or incompletely documented previously"[6]. For spontaneous reporting system, many machine learning methods are used to detect ADRs, such as Bayesian confidence propagation neural network (BCPNN) [7], decision support method [8], genetic algorithm [9], knowledge based approach [10], etc. One limitation is the reporting mechanism to submit ADR reports [8], which has serious underreporting and is not able to accurately quantify the corresponding risk. Another limitation is hard to detect ADRs with small number of occurrences of each drug-event association in the database.

In this paper we propose feature selection approach to detect ADRs from The Health Improvement Network (THIN) database. First feature matrix, which represents the medical events for the patients before and after taking drugs, is created by linking patients' prescriptions and corresponding medical events together. Then significant features are selected based on feature selection methods, comparing the feature matrix before patients take drugs with one after patients take drugs. Finally the significant ADRs can be detected from thousands of medical events based on corresponding features. Experiments are carried out on the drug Atorvastatin. Good performance is achieved.

## II. FEATURE MATRIX AND FEATURE SELECTION

### A. The Extraction of Feature Matrix

To detect the ADRs of drugs, first feature matrix is extracted from THIN database, which describes the medical events that patients occur before or after taking drugs. Then feature selection method of Student's t-test is performed to select the significant features from feature matrix containing thousands of medical events. Figure 1 shows the process to detect the ADRs using feature matrix. Feature matrix *A* describes the medical events for each patient during 60 days before they take drugs. Feature matrix *B* reflects the medical events during 60 days after patients take drugs. In order to reduce the effect of the small events, and save the computation time and space, we set 100 patients as a group. Matrix *X* and *Y* are feature matrix after patients are divided into groups.

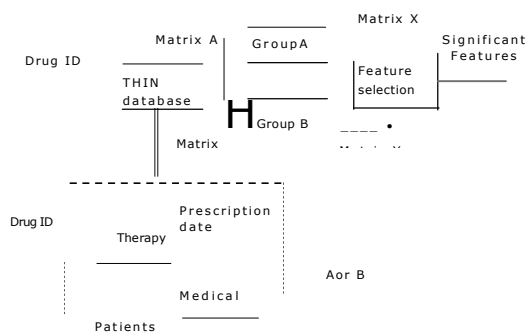

Fig.1. The process to detect ADRs. Matrix *A* and *B* are feature matrix before patients take drugs or after patients take drugs. The time period of observation is set to 60 days. Matrix *X* and *Y* are feature matrix after patients are divided into groups. We set 100 patients as one group.

### B. Medical Events and Readcodes

Medical events or symptoms are represented by medical codes or Readcodes. There are 103387 types of medical events in "Readcodes" database. The Read Codes used in general practice (GP), were invented and developed by Dr James Read in 1982. The NHS (National Health Service) has expanded the codes to cover all areas of clinical practice. The code is hierarchical from left to right or from level 1 to level 5. It means that it gives more detailed information from level 1 to level 5. Table 1 shows the medical symptoms based on Readcodes at level 3 and at level 5. 'Other soft tissue disorders' is general term using Readcodes at level 3. `Foot pain', 'Heel pain', etc., give more details using Readcodes at level 5.

## C. Feature Selection Based on Student's t-test

Feature extraction and feature selection are widely used in biomedical data processing [11-17]. In our research we use Student's t-test [18] feature selection method to detect the significant ADRs from thousands of medical events. Student's t-test is a kind of statistical hypothesis test based on a normal distribution, and is used to measure the difference between two kinds of samples. \

TABLE I MEDICAL EVENTS BASED ON READCODES AT LEVEL 3 AND LEVEL 5.

|  | Level | Readcodes | Medical events |
|---|---|---|---|
| Muscle pain | Level 3 | N24..00 | Other soft tissue disorders |
|  | Level 5 | N245.16 | Leg pain |
|  |  | N245111 | Toe pain |
|  |  | N245.13 | Foot pain |
|  |  | N245700 | Shoulder pain |
|  |  | N245.15 | Heel pain |

## D. Other Parameters

The variable of ratio $R_1$ is defined to evaluate significant changes of the medical events, using ratio of the patient number after taking the drug to one before taking the drug. The variable $R_2$ represents the ratio of patient number after taking the drug to the number of whole population having one particular medical symptom.

The ratio variables $R_1$ and $R_2$ are defined as follows:

$$R_1 = \frac{N_A}{N_B} \quad if\ N_B \neq 0; \quad (2)$$

$$\quad = N_A \quad if\ N_B = 0;$$

$$R_2 = N_A / N$$

where $N_B$ and $N_A$ represent the numbers of patients before or after they take drugs for having one particular medical event respectively. The variable $N$ represents the number of whole population who take drugs.

## III. EXPERIMENTS AND RESULTS

Atorvastatin [19], under the trade name Lipitor, is one of the drugs known as statins. It is used for lowering blood cholesterol. Drugs.com provides access to healthcare information tailored for a professional audience, sourced solely from the most trusted, well-respected and independant agents such as the Food and Drug Administration (FDA), American Society of Health-System Pharmacists, Wolters Kluwer Health, Thomson Micromedex, Cerner Multum and Stedman's. The side effects for Atorvastatin [20] include dark urine, muscle pain, tenderness, or weakness, painful, frequent urination, redness, or swelling of a tendon or joint, swollen, blistered, vomiting, stomach pain, unusual tiredness, and yellowing of the eyes or skin, etc.

6803 patients are taking Atorvastatin from 20GPs data in THIN database. Based on Readcodes at level 1-5, totally 10528 medical events are obtained before or after 6803 patients take the drug. So 6803x10528 feature matrix is obtained. Based on Readcodes at level 1-3, we combine the medical events, which have the same first three codes, into one medical event. Totally 2350 medical events are obtained, and 6803x2350 feature matrix are created. After grouping them, 68x10528 and 68x2350 feature matrix are formed to select the significant features, which reflect the significant change of medical events.

Table 2 shows the top 20 detected results in ascending order of p value of Student's t-test, using Readcodes at level 1-5 and at level 1-3. The detected results are using p value less than 0.05, which represent the significant change after patients take the drug. Table 3 shows the results in descending order of the ratio of the number of patients after taking the drug to one before taking the drug. Table 4 shows cancer information related to Atorvastatin. The detected ADRs are based on our computerized method, further investigation is needed.

It is clear that our detected results are consistent with published side effects for statin drugs [20, 21]. Major ADRs of 'muscle and musculoskeletal' events for statin drugs are detected not only based on Readcodes at level 1-5, but also based on Readcodes at level 1-3.

## IV. CONCLUSIONS

In this study we propose a novel method to successfully detect the ADRs using feature matrix and feature selection. A feature matrix, which characterizes the medical events before patients take drugs or after patients take drugs, is created from THIN database. The feature selection method of Student's t-test is used to detect the significant features from thousands of medical events. The significant ADRs, which are corresponding to significant features, are detected. Experiments are performed on the drug Atorvastatin. Compared to other computerized method, our proposed method achieves good performance.

TABLE II. THE TOP 20 ADRs FOR ATORVASTATIN BASED ON P VALUE OF STUDENT'S T-TEST.

| | Rank | Readcodes | Medical events | NB | NA | R1 | R2 |
|---|---|---|---|---|---|---|---|
| Level 1-5 | 1 | 1A55.00 | Dysuria | 26 | 181 | 6.96 | 2.66 |
| | 2 | 1Z12.00 | Chronic kidney disease stage 3 | 63 | 477 | 7.57 | 7.01 |
| | 3 | N131.00 | Cervicalgia - pain in neck | 73 | 337 | 4.62 | 4.95 |
| | 4 | N143.00 | Sciatica | 48 | 205 | 4.27 | 3.01 |
| | 5 | F4C0.00 | Acute conjunctivitis | 60 | 274 | 4.57 | 4.03 |
| | 6 | N245.17 | Shoulder pain | 99 | 376 | 3.80 | 5.53 |
| | 7 | M03z000 | Cellulitis NOS | 57 | 264 | 4.63 | 3.88 |
| | 8 | 1C14.00 | "Blocked ear" | 13 | 120 | 9.23 | 1.76 |
| | 9 | A53..11 | Shingles | 17 | 121 | 7.12 | 1.78 |
| | 10 | H01..00 | Acute sinusitis | 44 | 194 | 4.41 | 2.85 |
| | 11 | 1M10.00 | Knee pain | 103 | 369 | 3.58 | 5.42 |
| | 12 | 19EA.00 | Change in bowel habit | 17 | 102 | 6.00 | 1.50 |
| | 13 | H06z000 | Chest infection NOS | 136 | 563 | 4.14 | 8.28 |
| | 14 | Cl OF.00 | Type 2 diabetes mellitus | 154 | 436 | 2.83 | 6.41 |
| | 15 | 1D14.00 | C/O: a rash | 70 | 357 | 5.10 | 5.25 |
| | 16 | M0...00 | Skin and subcutaneous tissue infections | 18 | 115 | 6.39 | 1.69 |
| | 17 | N247100 | Leg cramps | 36 | 132 | 3.67 | 1.94 |
| | 18 | M111.00 | Atopic dermatitis/eczema | 36 | 190 | 5.28 | 2.79 |
| | 19 | 19F..00 | Diarrhoea symptoms | 52 | 244 | 4.69 | 3.59 |
| | 20 | H33..00 | Asthma | 23 | 106 | 4.61 | 1.56 |
| Level 1-3 | 1 | 171..00 | Cough | 309 | 1105 | 3.58 | 16.24 |
| | 2 | N21..00 | Peripheral enthesopathies and allied | 136 | 568 | 4.18 | 8.35 |
| | 3 | N24..00 | Other soft tissue disorders | 462 | 1371 | 2.97 | 20.15 |
| | 4 | 1Z1..00 | Chronic renal impairment | 79 | 570 | 7.22 | 8.38 |
| | 5 | 1B1..00 | General nervous symptoms | 204 | 667 | 3.27 | 9.80 |
| | 6 | 19F..00 | Diarrhoea symptoms | 97 | 444 | 4.58 | 6.53 |
| | 7 | M22..00 | Other dermatoses | 94 | 353 | 3.76 | 5.19 |
| | 8 | 1C1..00 | Hearing symptoms | 43 | 269 | 6.26 | 3.95 |
| | 9 | 183..00 | Oedema | 97 | 433 | 4.46 | 6.36 |
| | 10 | N13..00 | Other cervical disorders | 81 | 362 | 4.47 | 5.32 |
| | 11 | 1B8..00 | Eye symptoms | 73 | 322 | 4.41 | 4.73 |
| | 12 | F4C..00 | Disorders of conjunctiva | 85 | 381 | 4.48 | 5.60 |
| | 13 | H06..00 | Acute bronchitis and bronchiolitis | 357 | 1161 | 3.25 | 17.07 |
| | 14 | 173..00 | Breathlessness | 198 | 675 | 3.41 | 9.92 |
| | 15 | 1A5..00 | Genitourinary pain | 56 | 276 | 4.93 | 4.06 |
| | 16 | M03..00 | Other cellulitis and abscess | 73 | 357 | 4.89 | 5.25 |
| | 17 | N09..00 | Other and unspecified joint disorders | 196 | 721 | 3.68 | 10.60 |
| | 18 | 1D1..00 | C/O: a general symptom | 159 | 695 | 4.37 | 10.22 |
| | 19 | A53..00 | Herpes zoster | 20 | 158 | 7.90 | 2.32 |
| | 20 | J57..00 | Other disorders of intestine | 48 | 222 | 4.63 | 3.26 |

Variable NB and NA represent the numbers of patients before or after they take drugs for having one particular medical event. Variable $R_1$ represents the ratio of the numbers of patients after taking drugs to the numbers of patients before taking drugs. Variable $R_2$ represents the ratio of the numbers of patients after taking drugs to the number of the whole population.

TABLE III. THE TOP 20 ADRS FOR ATORVASTATIN BASED ON DESCENDING ORDER OF R1 VALUE.

| | Rank | Readcodes | Medical events | NB | NA | R1 | R2 |
|---|---|---|---|---|---|---|---|
| Level 1-5 | 1 | M03z100 | Abscess NOS | 1 | 39 | 39.00 | 0.57 |
| | 2 | 1922.00 | Sore mouth | 1 | 29 | 29.00 | 0.43 |
| | 3 | 173K.00 | MRC Breathlessness Scale: grade 4 | 1 | 28 | 28.00 | 0.41 |
| | 4 | F420600 | Non proliferative diabetic retinopathy | 1 | 28 | 28.00 | 0.41 |
| | 5 | M22z.11 | Actinic keratosis | 1 | 27 | 27.00 | 0.40 |
| | 6 | 564..13 | Head injury | 1 | 24 | 24.00 | 0.35 |
| | 7 | Sz...00 | Injury and poisoning NOS | 1 | 24 | 24.00 | 0.35 |

|  | | Readcodes | | Medical events | NB | NA | R1 |
|---|---|---|---|---|---|---|---|
| | 8 | G843.00 | External haemorrhoids, simple | 0 | 22 | 22.00 | 0.32 |
| | 9 | J570100 | Rectal polyp | 1 | 22 | 22.00 | 0.32 |
| | 10 | C10FL00 | Type 2 diabetes mellitus with persistent proteinuria | 0 | 22 | 22.00 | 0.32 |
| | 11 | G5yy900 | Left ventricular systolic dysfunction | 1 | 22 | 22.00 | 0.32 |
| | 12 | J681.00 | Melaena | 0 | 21 | 21.00 | 0.31 |
| | 13 | 11060.11 | Acute wheezy bronchitis | 1 | 21 | 21.00 | 0.31 |
| | 14 | N141.00 | Pain in thoracic spine | 1 | 21 | 21.00 | 0.31 |
| | 15 | N216500 | Prepatellar bursitis | 1 | 20 | 20.00 | 0.29 |
| | 16 | F4C0311 | Sticky eye | 1 | 20 | 20.00 | 0.29 |
| | 17 | 1M00.00 | Pain in elbow | 1 | 20 | 20.00 | 0.29 |
| | 18 | N211300 | Supraspinatus tendinitis | 1 | 20 | 20.00 | 0.29 |
| | 19 | J680.00 | Haematemesis | 1 | 19 | 19.00 | 0.28 |
| | 20 | F425.00 | Degeneration of macula and posterior pole | 1 | 19 | 19.00 | 0.28 |
| Level 1-3 | 1 | 1J0..00 | Suspected malignancy | 0 | 28 | 28.00 | 0.41 |
| | 2 | BB2..00 | [M]Papillary and squamous cell neoplasms | 0 | 24 | 24.00 | 0.35 |
| | 3 | 1M0..00 | Pain in upper limb | 1 | 24 | 24.00 | 0.35 |
| | 4 | S3z..00 | Fracture of unspecified bones | 1 | 24 | 24.00 | 0.35 |
| | 5 | Sz...00 | Injury and poisoning NOS | 1 | 24 | 24.00 | 0.35 |
| | 6 | J68..00 | Gastrointestinal haemorrhage | 3 | 54 | 18.00 | 0.79 |
| | 7 | SD9..00 | Superficial injuries of multiple and unspecified sites | 0 | 18 | 18.00 | 0.26 |
| | 8 | J04..00 | Dentofacial anomalies | 1 | 17 | 17.00 | 0.25 |
| | 9 | D01..00 | Other deficiency anaemias | 2 | 33 | 16.50 | 0.49 |
| | 10 | J07..00 | Salivary gland diseases | 1 | 16 | 16.00 | 0.24 |
| | 11 | C35..00 | Disorders of mineral metabolism | 1 | 16 | 16.00 | 0.24 |
| | 12 | M08..00 | Cutaneous cellulitis | 2 | 29 | 14.50 | 0.43 |
| | 13 | G86..00 | Noninfective lymphatic disorders | 0 | 14 | 14.00 | 0.21 |
| | 14 | J03..00 | Gingival and periodontal disease | 1 | 14 | 14.00 | 0.21 |
| | 15 | 172..00 | Blood in sputum - haemoptysis | 2 | 26 | 13.00 | 0.38 |
| | 16 | PI13..00 | Other specified skin anomalies | 0 | 12 | 12.00 | 0.18 |
| | 17 | 535..00 | Fracture of one or more tarsal and metatarsal bones | 0 | 12 | 12.00 | 0.18 |
| | 18 | J67..00 | Diseases of pancreas | 1 | 12 | 12.00 | 0.18 |
| | 19 | SD7..00 | Superficial injury of foot and toe(s) | 1 | 12 | 12.00 | 0.18 |
| | 20 | D31..00 | Purpura and other haemorrhagic conditions | 1 | 12 | 12.00 | 0.18 |

TABLE W. THE POTENTIAL ADRS RELATED TO CANCER FOR ATORVASTATIN BASED ON P VALUE OF STUDENT'S T-TEST.

| | Rank | Readcodes | | Medical events | NB | NA | R1 |
|---|---|---|---|---|---|---|---|
| 1 | B33..00 | Other malignant neoplasm of skin | | 18 | 97 | 5.39 | 1.43 |
| 2 | B76..00 | Benign neoplasm of skin | | 52 | 139 | 2.67 | 2.04 |
| 3 | 1J0..00 | Suspected malignancy | | 0 | 28 | 28.00 | 0.41 |
| 4 | B46..00 | Malignant neoplasm of prostate | | 8 | 46 | 5.75 | 0.68 |
| 5 | BB2..00 | [M]Papillary and squamous cell neoplasms | | 0 | 24 | 24.00 | 0.35 |
| 6 | B34..00 | Malignant neoplasm of female breast | | 4 | 26 | 6.50 | 0.38 |
| 7 | BB5..00 | [M]Adenomas and adenocarcinomas | | 6 | 28 | 4.67 | 0.41 |
| 8 | B13..00 | Malignant neoplasm of colon | | 3 | 20 | 6.67 | 0.29 |
| 9 | B14..00 | Malignant neoplasm of rectum, rectosigmoid junction and anus | | 0 | 9 | 9.00 | 0.13 |
| 10 | B8..00 | Carcinoma in situ | | 3 | 19 | 6.33 | 0.28 |
| 11 | B83..00 | Carcinoma in situ of breast and genitourinary system | | 7 | 21 | 3.00 | 0.31 |
| 12 | BB4..00 | [M]Transitional cell papillomas and carcinomas | | 0 | 7 | 7.00 | 0.10 |
| 13 | B49..00 | Malignant neoplasm of urinary bladder | | 2 | 11 | 5.50 | 0.16 |
| 14 | B22..00 | Malignant neoplasm of trachea, bronchus and lung | | 2 | 15 | 7.50 | 0.22 |
| 15 | B10..00 | Malignant neoplasm of oesophagus | | 1 | 10 | 10.00 | 0.15 |
| 16 | B57..00 | Secondary malig neop of respiratory and digestive systems | | 0 | 6 | 6.00 | 0.09 |
| 17 | BB3..00 | [M]Basal cell neoplasms | | 4 | 19 | 4.75 | 0.28 |
| 18 | B58..00 | Secondary malignant neoplasm of other specified sites | | 0 | 9 | 9.00 | 0.13 |
| 19 | B59..00 | Malignant neoplasm of unspecified site | | 0 | 5 | 5.00 | 0.07 |
| 20 | B21..00 | Malignant neoplasm of larynx | | 0 | 5 | 5.00 | 0.07 |
| 21 | BB8..00 | [M]Cystic, mucinous and serous neoplasms | | 0 | 5 | 5.00 | 0.07 |
| 22 | B11..00 | Malignant neoplasm of stomach | | 1 | 7 | 7.00 | 0.10 |
| 23 | B4A..00 | Malig neop of kidney and other unspecified urinary organs | | 0 | 6 | 6.00 | 0.09 |
| 24 | B7F..00 | Benign neoplasm of brain and other parts of nervous system | | 0 | 6 | 6.00 | 0.09 |
| 25 | BB1..00 | [M]Epithelial neoplasms NOS | | 2 | 10 | 5.00 | 0.15 |
| 26 | B71..00 | Benign neoplasm of other parts of digestive system | | 4 | 11 | 2.75 | 0.16 |
| 27 | B812.00 | Carcinoma in situ of bronchus and lung | | 0 | 2 | 2.00 | 0.03 |
| 28 | BBK..00 | [M]Myomatous neoplasms | | 0 | 2 | 2.00 | 0.03 |
| 29 | B440.00 | Malignant neoplasm of ovary | | 0 | 2 | 2.00 | 0.03 |
| 30 | BBMz.00 | [M]Fibroepithelial neoplasm NOS | | 0 | 2 | 2.00 | 0.03 |